\documentclass{webofc}
\usepackage[varg]{txfonts}   % Web of Conferences font
\usepackage{color}

\begin{document}
\title{Searching for high-z DSFGs with NIKA2 and NOEMA}
\author{\firstname{L.}~\lastname{Bing}\inst{\ref{LAM}}\fnsep\thanks{\email{longji.bing@lam.fr}}
  \and \firstname{R.}~\lastname{Adam} \inst{\ref{LLR}}
  \and  \firstname{P.}~\lastname{Ade} \inst{\ref{Cardiff}}
  \and  \firstname{H.}~\lastname{Ajeddig} \inst{\ref{CEA}}
  \and  \firstname{P.}~\lastname{Andr\'e} \inst{\ref{CEA}}
  \and \firstname{E.}~\lastname{Artis} \inst{\ref{LPSC}}
  \and  \firstname{H.}~\lastname{Aussel} \inst{\ref{CEA}}
  \and  \firstname{A.}~\lastname{Beelen} \inst{\ref{IAS}}
  \and  \firstname{A.}~\lastname{Beno\^it} \inst{\ref{Neel}}
  \and  \firstname{S.}~\lastname{Berta} \inst{\ref{IRAMF}}
  \and  \firstname{M.}~\lastname{B\'ethermin} \inst{\ref{LAM}}
  \and  \firstname{O.}~\lastname{Bourrion} \inst{\ref{LPSC}}
  \and  \firstname{M.}~\lastname{Calvo} \inst{\ref{Neel}}
  \and  \firstname{A.}~\lastname{Catalano} \inst{\ref{LPSC}}
  \and  \firstname{M.}~\lastname{De~Petris} \inst{\ref{Roma}}
  \and  \firstname{F.-X.}~\lastname{D\'esert} \inst{\ref{IPAG}}
  \and  \firstname{S.}~\lastname{Doyle} \inst{\ref{Cardiff}}
  \and  \firstname{E.~F.~C.}~\lastname{Driessen} \inst{\ref{IRAMF}}
  \and  \firstname{A.}~\lastname{Gomez} \inst{\ref{CAB}}
  \and  \firstname{J.}~\lastname{Goupy} \inst{\ref{Neel}}
  \and  \firstname{F.}~\lastname{K\'eruzor\'e} \inst{\ref{LPSC}}
  \and  \firstname{C.}~\lastname{Kramer} \inst{\ref{IRAME}}
  \and  \firstname{B.}~\lastname{Ladjelate} \inst{\ref{IRAME}}
  \and  \firstname{G.}~\lastname{Lagache} \inst{\ref{LAM}}
  \and  \firstname{S.}~\lastname{Leclercq} \inst{\ref{IRAMF}}
  \and  \firstname{J.-F.}~\lastname{Lestrade} \inst{\ref{LERMA}}
  \and  \firstname{J.-F.}~\lastname{Mac\'ias-P\'erez} \inst{\ref{LPSC}}
  \and  \firstname{A.}~\lastname{Maury} \inst{\ref{CEA}}
  \and  \firstname{P.}~\lastname{Mauskopf} \inst{\ref{Cardiff},\ref{Arizona}}
  \and \firstname{F.}~\lastname{Mayet} \inst{\ref{LPSC}}
  \and  \firstname{A.}~\lastname{Monfardini} \inst{\ref{Neel}}
  \and  \firstname{M.}~\lastname{Mu\~noz-Echeverr\'ia} \inst{\ref{LPSC}}
  \and  \firstname{R.}~\lastname{Neri} \inst{\ref{IRAMF}}
  \and  \firstname{A.}~\lastname{Omont} \inst{\ref{IAP}}
  \and  \firstname{L.}~\lastname{Perotto} \inst{\ref{LPSC}}
  \and  \firstname{G.}~\lastname{Pisano} \inst{\ref{Cardiff}}
  \and  \firstname{N.}~\lastname{Ponthieu} \inst{\ref{IPAG}}
  \and  \firstname{V.}~\lastname{Rev\'eret} \inst{\ref{CEA}}
  \and  \firstname{A.~J.}~\lastname{Rigby} \inst{\ref{Cardiff}}
  \and  \firstname{A.}~\lastname{Ritacco} \inst{\ref{IAS}, \ref{ENS}}
  \and  \firstname{C.}~\lastname{Romero} \inst{\ref{Pennsylvanie}}
  \and  \firstname{H.}~\lastname{Roussel} \inst{\ref{IAP}}
  \and  \firstname{F.}~\lastname{Ruppin} \inst{\ref{MIT}}
  \and  \firstname{K.}~\lastname{Schuster} \inst{\ref{IRAMF}}
  \and  \firstname{S.}~\lastname{Shu} \inst{\ref{Caltech}}
  \and  \firstname{A.}~\lastname{Sievers} \inst{\ref{IRAME}}
  \and  \firstname{C.}~\lastname{Tucker} \inst{\ref{Cardiff}}
  \and  \firstname{R.}~\lastname{Zylka} \inst{\ref{IRAMF}}}
  
  \institute{
    Univ. Grenoble Alpes, CNRS, Grenoble INP, LPSC-IN2P3, 53, avenue des Martyrs, 38000 Grenoble, France
    \label{LPSC}
    \and
    Univ. Grenoble Alpes, CNRS, IPAG, 38000 Grenoble, France 
    \label{IPAG}
    \and
    LLR (Laboratoire Leprince-Ringuet), CNRS, École Polytechnique, Institut Polytechnique de Paris, Palaiseau, France
    \label{LLR}
    \and
    School of Physics and Astronomy, Cardiff University, Queen’s Buildings, The Parade, Cardiff, CF24 3AA, UK 
    \label{Cardiff}
    \and
    AIM, CEA, CNRS, Universit\'e Paris-Saclay, Universit\'e Paris Diderot, Sorbonne Paris Cit\'e, 91191 Gif-sur-Yvette, France
    \label{CEA}
    \and
    Institut d'Astrophysique Spatiale (IAS), CNRS, Universit\'e Paris Sud, Orsay, France
    \label{IAS}
    \and
    Institut N\'eel, CNRS, Universit\'e Grenoble Alpes, France
    \label{Neel}
    \and
    Institut de RadioAstronomie Millim\'etrique (IRAM), Grenoble, France
    \label{IRAMF}
    \and 
    Dipartimento di Fisica, Sapienza Universit\`a di Roma, Piazzale Aldo Moro 5, I-00185 Roma, Italy
    \label{Roma}
    \and
    Centro de Astrobiolog\'ia (CSIC-INTA), Torrej\'on de Ardoz, 28850 Madrid, Spain
    \label{CAB}
    \and  
    Instituto de Radioastronom\'ia Milim\'etrica (IRAM), Granada, Spain
    \label{IRAME}
    \and
    Aix Marseille Univ, CNRS, CNES, LAM (Laboratoire d'Astrophysique de Marseille), Marseille, France
    \label{LAM}
    \and 
    LERMA, Observatoire de Paris, PSL Research University, CNRS, Sorbonne Universit\'e, UPMC, 75014 Paris, France  
    \label{LERMA}
    \and 
    Institut d'Astrophysique de Paris, CNRS (UMR7095), 98 bis boulevard Arago, 75014 Paris, France
    \label{IAP}
    \and
    Department of Physics and Astronomy, University of Pennsylvania, 209 South 33rd Street, Philadelphia, PA, 19104, USA
    \label{Pennsylvanie}
    Laboratoire de Physique de l’\'Ecole Normale Sup\'erieure, ENS, PSL Research University, CNRS, Sorbonne Universit\'e, Universit\'e de Paris, 75005 Paris, France 
    \label{ENS}
    Kavli Institute for Astrophysics and Space Research, Massachusetts Institute of Technology, Cambridge, MA 02139, USA
    \label{MIT}
    \and
    School of Earth and Space Exploration and Department of Physics, Arizona State University, Tempe, AZ 85287, USA
    \label{Arizona}
    \and
    Caltech, Pasadena, CA 91125, USA
    \label{Caltech}
  }
  
  % In addition to the (non)detections of lines in the spectra, the total IR luminosities estimated by the SED fitting are also used to predict the line fluxes and evaluate if the (non)detections are consistent with the expected line fluxes at given redshifts. 
  
 \abstract{%
As the possible progenitors of passive galaxies at z=2-3, dusty star-forming galaxies (DSFGs) at z>4 provide a unique perspective to study the formation, assembly, and early quenching of massive galaxies in the early Universe. The extreme obscuration in optical-IR makes (sub)mm spectral scans the most universal and unbiased way to confirm/exclude the high-z nature of candidate dusty star-forming galaxies. We present here the status of the NIKA2 Cosmological Legacy Survey (N2CLS), which is the deepest wide-area single-dish survey in the millimeter searching for high-z DSFGs. We also introduce a joint-analysis method to efficiently search for the spectroscopic redshift of high-z DSFGs with noisy spectra and photometric data and present its success in identifying the redshift of DSFGs found in NIKA2 science verification data.
}
\maketitle
\section{Introduction}
\label{intro}
In the past decades, optical-IR deep-field observations have contributed to the studies on galaxy assembly in the early Universe \cite{MD14}. The rest-frame UV-optical emission from young stars is widely used to constrain the cosmic star formation rate density (SFRD) \cite{Bouwens14}, one of the key quantities to calibrate cosmological simulations and galaxy evolution models \cite{Pillepich18}. However, UV emission is also heavily affected by the extinction caused by dusty ISM. Dust extinction becomes increasingly prevalent in massive high-z galaxies, making the majority of them almost invisible even in near-IR. Thus, the SFRD and space density of massive galaxies might be underestimated in studies with only optical and near-IR data \cite{Wang18}.

While most of the UV photons are absorbed, their energy heat the dusty ISM and is re-emitted mostly in far-IR. This makes dusty star-forming galaxies (DSFGs) bright in far-IR to the millimeter and constitutes the majority of sources in blind ALMA and single-dish observations. Complementary to optical and near-IR observations, far-IR to millimeter surveys provide additional constraints on total SFRD and the rise of massive galaxies at z$<$3 \cite{Dunlop17}. However, SFRDs measured at z$>$4 still differ from each other by more than a factor of ten \cite{Gruppioni20, D20, Zavala21}. This is mainly due to the difficulties in defining a large unbiased high-z DSFG sample and properly reconstructing the infrared luminosity function (IRLF). Current ALMA surveys \cite{Gruppioni20, Franco18} uncover a few blindly detected $z>4$ sources. However, their small area limits the sample size, hence the bright end of the IRLF is
still poorly sampled.

%With growing contribution to the full population at higher redshift\cite{Gruppioni13}, these luminous sources are increasingly crucial to study the evolution of IRLF and the obscured SFRD at z$>$4.

To efficiently uncover a large sample of these rare sources, large area blind surveys are required. Nowadays, such surveys could only be achieved by large field-of-view continuum cameras on single-dish telescopes, like SCUBA-2 on Jame Clark Maxwell Telescope and NIKA2 on IRAM 30m \cite{NIKA2-performance, NIKA2-general, NIKA2-instrument, NIKA2-electronics}. However, the large beam size of these data and the high frequency of optical/radio dark sources make it difficult to identify their counterparts and measure their redshifts, infrared luminosity (L$_{IR}$), and other properties. NOEMA and ALMA interferometers provide the best resolution and sensitivity at mm wavelengths to measure the accurate positions and detect the spectral lines emitted by these sources.
 
 In this paper, we first present the current status of the NIKA2 Cosmological Legacy Survey (N2CLS), an ongoing NIKA2 guaranteed time observation (GTO) large program searching for large samples of high-z DSFGs. We also introduce a new method to search for the redshift of DSFGs combining broadband photometry and mm spectral scan.
 
 %Interferometers like ALMA and NOEMA provide the best resolution and sensitivity at mm to get the accurate positions and emission lines for source identification and redshift searching, while it is prevalent that blind spectral scans on sources failed to find unambiguous redshift solution, with less than 2 spectral line detections. 
 %\GL{Plan of the paper}
 
 %We will present the current status of NIKA2 Cosmological Legacy Survey(N2CLS), the deepest wide-area single-dish survey on the millimeter sky in Sect.~\ref{sec-1}. In Sect.~\ref{sec-2} we will introduce a joint-analysis method to constrain the redshift of DSFGs with NOEMA spectral scans and SED analysis, and test its feasibility on follow-up observations on NIKA2 detected DSFGs.

\section{Basic Information and Status of N2CLS}
\label{sec-1}
N2CLS is a NIKA2 GTO large program observing 2 deep fields in the northern/equatorial sky: GOODS-N and COSMOS. The observations are designed to have good statistics on both bright and faint sources. The survey in GOODS-N covers a relatively narrow area of 160 arcmin$^2$ but is designed to approach the confusion limit of the IRAM 30m telescope. The survey in COSMOS covers a much larger area of $\sim$1100 arcmin$^2$ with a shallower depth, to get a larger sample of bright sources and very high redshift DSFGs. These two fields benefit from the multi-wavelength coverage by deep observations from the X-ray to radio. The sources detected by NIKA2 are cross-matched to existing catalogs across the whole electromagnetic spectrum to identify their counterparts, derive their redshift, and study their physical properties.

We have already completed 85.2h observations on GOODS-N and 82.9h on COSMOS, by March 2021. The data are reduced by PIIC, the data reduction pipeline supported by IRAM\cite{Zylka13}, as well as the NIKA2 collaboration internal pipeline. Table.~\ref{tab-1} shows the current noise level in the central region of N2CLS fields compared to other (sub)mm surveys in the same fields. The data available so far in GOODS-N have already surpassed the depth of any other millimeter surveys performed with AzTEC \cite{azgn}, MAMBO \cite{mambogn} and GISMO \cite{gismogn}, and they match with the deepest SCUBA2 survey at higher frequency \cite{scuba2gn}. 62/25 sources are detected with SNR>4 at 1.2/2mm. As for COSMOS, the survey has also been the deepest single-dish millimeter survey compared to AzTEC  \cite{azcos}, MAMBO \cite{mambo} and GISMO \cite{gismocos}. The normalized depth is still slightly shallower than the S2COSMOS survey \cite{s2cosmos} at 850$\mu$m and more observations in the coming season in winter 2021-2022 are expected to make the 2 surveys comparable in depth. 157/63 sources are detected in COSMOS with SNR>4 at 1.2/2mm.

The N2CLS survey aims to produce a full census of high-z DSFGs in
these two fields. Combining the two fields, we expect to improve the current estimate on the millimeter number counts and provide new means to rule out or confirm state-of-the-art galaxy evolution models and simulations. In particular, combined with the wealth of ancillary data, N2CLS will put new solid constraints on the obscured SFRD at z>3. To reach that goal, we need first to obtain the redshift of N2CLS sources. The deep optical-IR data have been extensively used to obtain photometric redshift of sources in GOODS-N and COSMOS with novel spectral energy distribution (SED) fitting methods. However, the high dust attenuation makes some DSFGs completely dark in optical, making it difficult to get their redshift with these methods. The blind mm spectral scans will be the only solution to measure their spectroscopic redshift.%\GL{Need z => introduce next section}

\begin{table*}
\caption{Basic information of N2CLS and other single-dish (sub)mm surveys in GOODS-N and COSMOS. For each survey, the root mean square (RMS) noise is normalized to 1.2/2mm using the average IR SED of star-forming galaxies at z=3 \cite{B15}. For NIKA2, the RMS of each band is obtained with current data in hands.}
\label{tab-1}       % Give a unique label
% For LaTeX tables you can use
\begin{tabular}{llccc}
%\hline
\hline
Field & Survey & $\nu_{obs}$ (GHz) & Area (arcmin$^2$) & RMS (mJy/beam)\\
\hline 
GOODS-N & AzTEC & 273 & 245 & 1.06 \\
& SCUBA2 & 353 & $\sim$140 & $\sim$0.16 \\
& GISMO & 150 & 38 & 0.14 \\
& N2CLS & 255 \& 150 & 160 & 0.19 \& 0.057  \\
\hline
COSMOS & AzTEC-ASTE & 273 & 2592 & 1.26 \\
& MAMBO & 255 & 400 & 1.00 \\
& S2COSMOS & 353 & 5760 & $\sim$0.47  \\
& GISMO & 150 & 250 & 0.23 \\
& N2CLS & 255 \& 150 & 1100 & 0.55 \& 0.18 \\\hline%\hline
\end{tabular}
% Or use
%\vspace*{2cm}  % with the correct table height
\end{table*}

%\subsection{Subsection title}
%\label{sec-1-1}
%Don't forget to give each section, subsection, subsubsection, and
%paragraph a unique label (see Sect.~\ref{sec-1}).

\section{Searching for the Redshift of DSFGs Scanned by NOEMA}
\label{sec-2}
%Redshift of DSFGs is one of the basic information required for studies on the obscured SFRD and the early assembly of massive galaxies. Spectral scans made by millimeter interferometers could obtain more robust redshifts of DSFG population with less biases and incompleteness caused by dust obscuration, but long observation time needed for line detections compared to continuum, and the minimum number of line detections(2 lines at least) for having unambiguous redshift solution on one source could not always be satisfied. In this section, we introduce a new framework to find the spectroscopic redshift of DSFGs with prior information from both photometric analysis and IR  luminosity from the SED fitting. 

%We will describe the basic concept of this framework and present its application to sources in NIKA2 science verification observations, to prove its reliability in finding the true spectroscopic redshift from several ambiguous redshift solutions . 

%\subsection{Method Description and Evaluation}
%\label{sec-2-1}

Before the start of N2CLS, we made several science verification observations to test the on-sky performance of the NIKA2 instrument, including one centered on the z=5.243 lensed DSFG J0918+5142 \cite{Combes12}. The NIKA2 observation covers $\sim$175 arcmin$^2$ and reaches RMS of 0.45/0.11 mJy/beam at 1.2/2\,mm. The surveyed area overlaps with the SPIRE footprint of Herschel Lensing Survey (HLS) \cite{Egami10} but is almost not covered by deep optical/near-IR observations. In this field, we detected 4 sources (HLS-2,3,4 and 22) bright in millimeters and with cold IR SEDs consistent with z=3-7 DSFGs. However, the lack of optical/near-IR data makes it difficult to obtain their accurate photometric redshift. To study their properties in detail, we made NOEMA band1 spectral scan observations from $\sim$71GHz to $\sim$102GHz on these sources to search for their redshift at first. Additional follow-up observations were made later in band1 and band2 based on the initial analysis on HLS-22 and HLS-2+3, respectively. 

%Blind spectral scan observations could be used to measure the redshift of DSFGs by detections of molecular/atomic lines. 
At least 2 lines detected with high significance are needed to get an unambiguous redshift solution, while we only got no more than 1 line detection on each source in our first observations. To design efficient follow-up observations, we develop a new method to search for the redshift of DSFGs jointly with spectral scans and broadband SED fitting, which could derive the redshift of DSFGs with high robustness from noisy mm spectra in our first observations. 

The basic concept of this method is using the redshift probability distribution and IR luminosity from the SED fitting, in addition to mm spectra, to jointly constrain the redshift. With far-IR and (sub-)mm photometric data, we compute the best estimate of IR luminosity at any given redshift, L$_{IR}(z)$. With the L$_{IR}(z)$, we further predict the total luminosity of CO, [CI] and [CII] lines using the empirical L$_{IR}$-L$_{line}$ relations \cite{Liu15} and then generate the model spectra at the given redshifts with our code. All spectral lines have Gaussian profiles of 500km/s line widths (FWHM) and fluxes scaled from the predicted luminosity at given redshifts. The line widths are chosen to be the average among bright sub-mm galaxies \cite{Spilker14}, which compromise the robustness and significance of the line identification under the interference of narrow noise spikes and a wide variety of line widths in different sources (see Fig.~3). Then, We compare the observed spectra with the predicted model at any given redshift to derive the goodness of match using the chi-square, noted as $\chi_{spec}^2(z)$. Combining with the chi-square of the SED fitting at any redshift ($\chi_{SED}^2$), we derive the likelihood of source redshift with Eq.~\ref{MLEz}:
\begin{equation}
\label{MLEz}
\mathcal{L}_{joint}(z) \propto \exp\{-[\chi_{SED}^2(z)+\chi_{spec}^2(z)]/2\}
\end{equation}

%\begin{table*}
%\centering
%\caption{SPIRE and NIKA2 fluxes (in mJy) on 4 HLS sources. Non detections are in brackets with 1$\sigma$. }
%\label{tab-2}     
%\begin{tabular}{cccccc}
%\hline
%Source & F$_{SPIRE250}$ & F$_{SPIRE350}$ & F$_{SPIRE500}$ & F$_{NIKA2-1.2}$ & F$_{NIKA2-2.0}$ \\
%\hline 
% HLS-2 & (6.06) & (6.08) & 17.45$\pm$6.14 & 2.93$\pm$0.29 & 0.42$\pm$0.07 \\
% HLS-3 & (6.07) & (6.09) & (6.34) & 2.39$\pm$0.26 & 0.60$\pm$0.06 \\
% HLS-4 & (6.06) & 8.89$\pm$6.07 & 8.91$\pm$6.10 & 1.92$\pm$0.27 & 0.28$\pm$0.07 \\
% HLS-22 & (6.08) & (6.09) & (6.11) & 1.68$\pm$0.26 & 0.36$\pm$0.06 \\
%\end{tabular}
%\end{table*}

\begin{figure}[h]
\begin{center}
\includegraphics[scale=0.23]{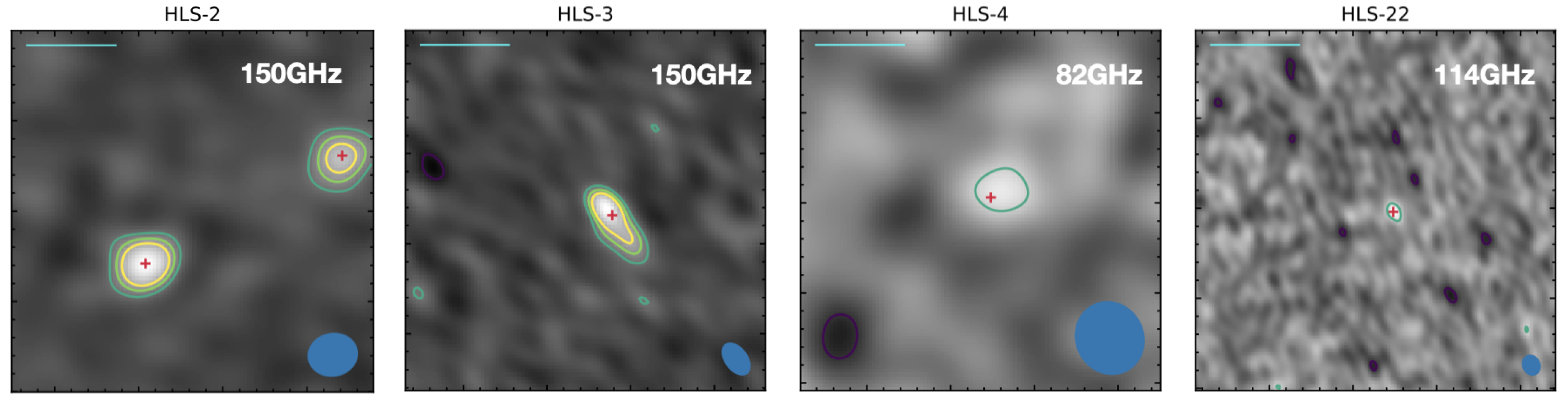}
\label{contimg}    
\caption{NOEMA continuum image of 4 sources in HLS field. Contour levels correspond to -3.5$\sigma$ (purple), 3.5$\sigma$, 5$\sigma$ and 6.5$\sigma$ (yellow), respectively. The NOEMA beams are shown in the bottom-right. }
\end{center}
\end{figure}
Fig.~1. shows the cleaned continuum image from NOEMA at the highest frequency setup. We make blind detection in these continuum uv-tables and keep sources with SNR>4 by NOEMA (red markers in Fig.~1). in the following study. Observation on HLS-2 reveals 2 sources with nearly equal fluxes with a separation of $\sim$10",  while the rest 3 do not show evidence of multiplicity. The spectra of the sources are then extracted from the full uv table, with the position and shape fixed to be the same as the continuum sources. 

We first applied the joint-analysis method to the spectra taken in 2017. 
We fit the SPIRE+NIKA2 SED with the SED templates of high-z star-forming galaxies from \cite{B15}, generate the model spectra, and derive the joint log-likelihood following the described procedures. Fig.~2. shows the log-likelihood of the redshift and spectral lines of HLS-2-1, HLS-3, and HLS-22 from the joint analysis and the SED only, with the maximum log-likelihood normalized to 1. HLS-2-1 has one marginally detected line at $\sim$92.3GHz, which matches with the discrete peaks in the normalized log-likelihood maximized at z=5.241. HLS-3 has no line detected in the band1 observation, and thus no significant peaks in log-likelihood. HLS-22 has one significant detection at $\sim$85.7 GHz, which also matches with the discrete peaks maximized at z=3.036 and z=4.380 in the log-likelihood. 

\begin{figure}[h]
\begin{center}
\includegraphics[scale=0.19]{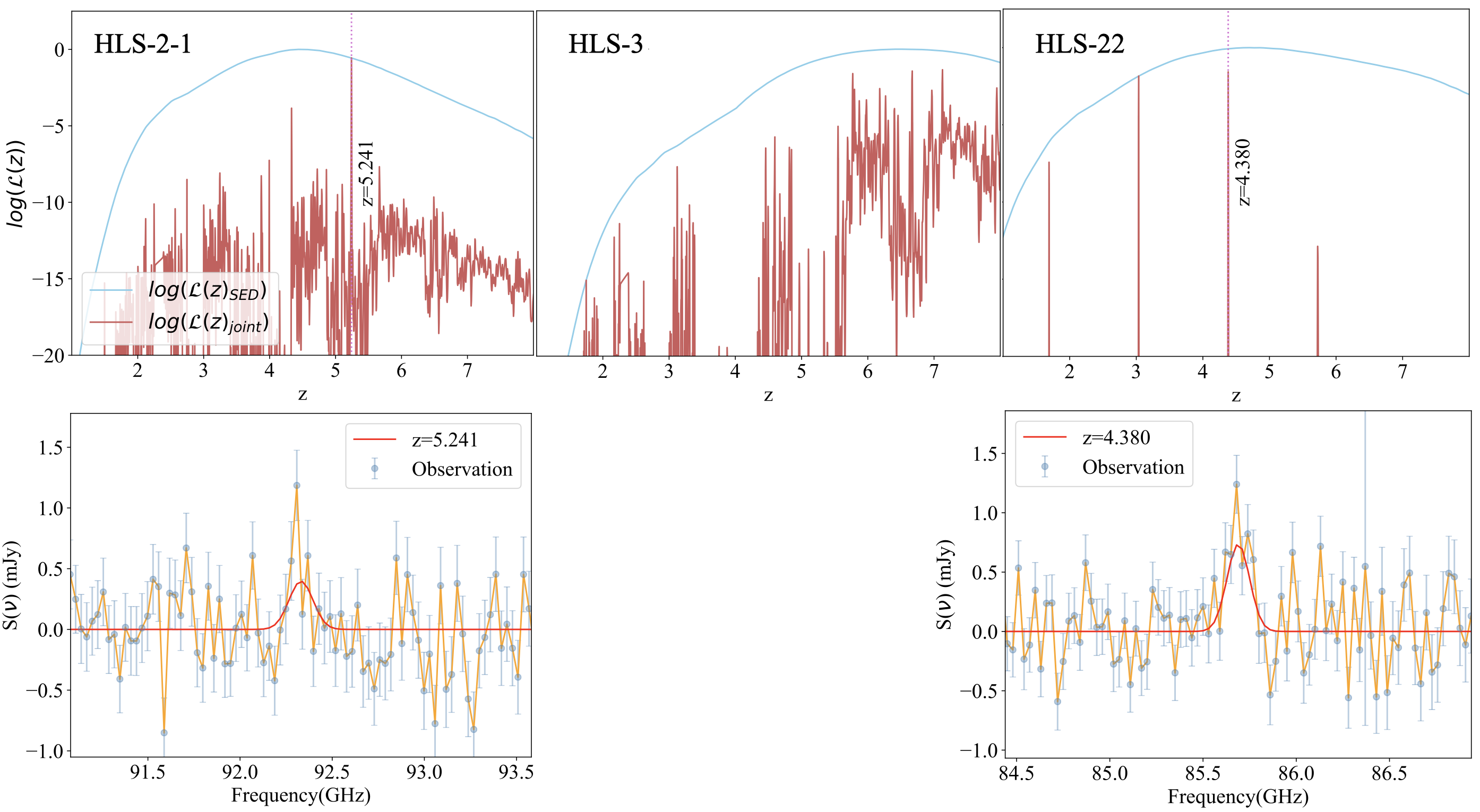}
\label{joint1}       % Give a unique label
\caption{Results of joint-analysis method with the extracted source spectra of 3 HLS sources observed in 2017. First row: The normalized log-likelihood of 3 HLS sources derived by our joint-analysis method. Second row: The blindly extracted source spectra and the model at the best redshift solution of HLS-2-1 and HLS-22 generated by the joint-analysis method.}
\end{center}
\end{figure}

We present the results of the joint analysis with further follow-up data in Fig.~3. For HLS-2-1, we confirm the redshift (z=5.241) in the previous analysis with a new line detection at 147.7\,GHz. With a new line detected at 114\,GHz, the redshift of HLS-22 is also confirmed to be at z=3.036, matching with one of the prominent peaks in the previous joint log-likelihood. As for HLS-3, the new analysis reveals one prominent peak in log-likelihood at z=3.123 matching to the line detected at 139.8\,GHz. This is not contradicting the previous results showing relatively high log-likelihood at this redshift under non-detection. %This suggests high noise levels of band1 spectral scan at the expected frequencies of emission lines compared to line fluxes, which results in low $\chi^2_{spec}$(z) even if there is no line detected. 
The consistency of the results from our joint-analysis method indicates the promising potential of applying it to the search of the redshift of DSFGs observed with blind millimeter spectral scans.

\begin{figure}[h]
\begin{center}
\includegraphics[scale=0.425]{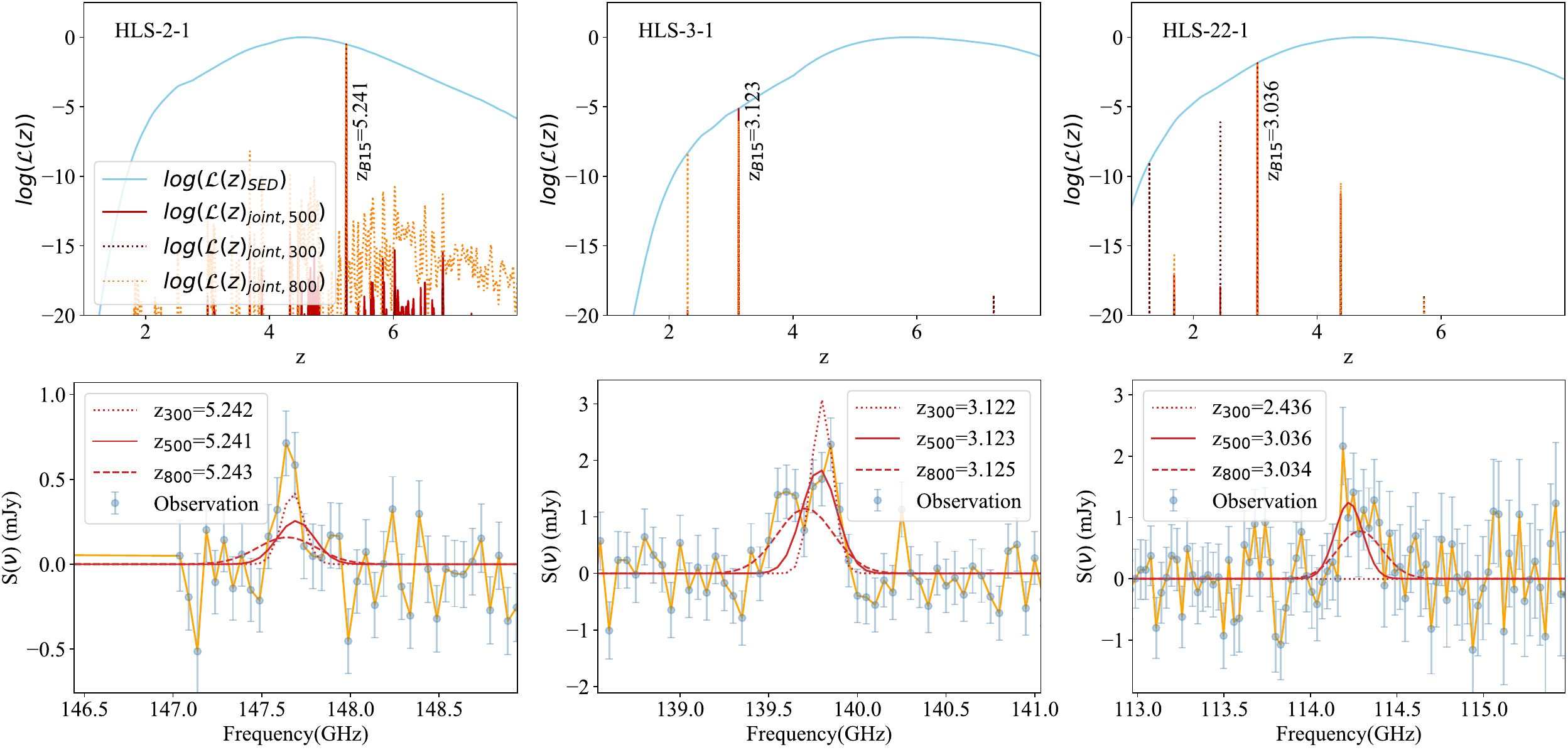}
\label{joint2}       % Give a unique label
\caption{Similar to Fig.~2, the results of the joint-analysis with new observations added and the spectra plus the model at the redshift with the maximum log-likelihood. Here we also present the results under FWHM 300/800 km/s for comparison. Except for the 300km/s model on HLS-22 affected by noise spikes, the redshift solutions remain consistent with those using the default 500 km/s model.}
\end{center}
\end{figure}

%For two-column wide figures use syntax of figure~\ref{fig-2}
%\begin{figure*}
%\centering
%\vspace*{5cm}       % Give the correct figure height in cm
%\label{fig-2}       % Give a unique label
%\end{figure*}

%For figure with sidecaption legend use syntax of figure
%\begin{figure}
%\centering
%\sidecaption
%\includegraphics[width=5cm,clip]{tiger}
%\label{fig-3}       % Give a unique label
%\end{figure}

%For tables use syntax in table~\ref{tab-1}.

\section*{Acknowledgements}
We would like to thank the IRAM staff for their support during the campaigns. The NIKA2 dilution cryostat has been designed and built at the Institut N\'eel. In particular, we acknowledge the crucial contribution of the Cryogenics Group, and in particular Gregory Garde, Henri Rodenas, Jean Paul Leggeri, Philippe Camus. This work has been partially funded by the Foundation Nanoscience Grenoble and the LabEx FOCUS ANR-11-LABX-0013. This work is supported by the French National Research Agency under the contracts "MKIDS", "NIKA" and ANR-15-CE31-0017 and in the framework of the "Investissements d’avenir” program (ANR-15-IDEX-02). This work has benefited from the support of the European Research Council Advanced Grant ORISTARS under the European Union's Seventh Framework Programme (Grant Agreement no. 291294). F.R. acknowledges financial supports provided by NASA through SAO Award Number SV2-82023 issued by the Chandra X-Ray Observatory Center, which is operated by the Smithsonian Astrophysical Observatory for and on behalf of NASA under contract NAS8-03060.

%
% BibTeX or Biber users please use (the style is already called in the class, ensure that the "woc.bst" style is in your local directory)
% \bibliography{name or your bibliography database}

\end{document}